\documentclass[12pt]{article}
\usepackage{epsfig,subfigure}
\usepackage{amssymb,amsmath,cite}

\newcommand{\ff}{ f_{f}} 

\begin{document}

\begin{titlepage}
\begin{flushright}
UCB-PTH-05/37\\
LBNL-59061\\
BUHEP-05-16\\
\end{flushright}

\vspace{15pt}

\begin{center}

{\huge\bf {Tensor Mesons in AdS/QCD}}

\vspace{15pt}

Emanuel Katz$^a$\footnote{amikatz@buphy.bu.edu},
Adam Lewandowski$^{a,b}$\footnote{lewandow@usna.edu}, and
Matthew D. Schwartz$^c$\footnote{mdschwartz@lbl.gov}\\

\vspace{7pt}
{\small

${}^a$
{\it{Department of Physics,\\
Boston University, Boston, MA 02215, USA \\
}
}
\vspace{4pt}
${}^b$
{\it Department of Physics, United States Naval Academy, \\
Annapolis, Maryland 21403, USA \\}
\vspace{4pt}
${}^c$ 
{\it{Department of Physics, University of California, Berkeley,\\
      and\\
      Theoretical Physics Group, Lawrence Berkeley National Laboratory,\\
      Berkeley, CA 94720, USA\\}
}
}
\end{center}

\begin{abstract}
We explore tensor mesons in AdS/QCD focusing on 
$f_2 (1270 )$, the lightest spin-two resonance in QCD. 
We find that the $f_2$ mass and the partial width 
$\Gamma ( f_2 \rightarrow\gamma \gamma )$ are in very good agreement
with data. In fact, the dimensionless ratio of these two
quantities comes out within the current experimental bound. 
The result for this ratio depends only on $N_c$ and $N_f$, and the 
quark and glueball content of the operator responsible for the $f_2$;
more importantly, it does not depend on chiral symmetry breaking and so is both
independent of much of the arbitrariness of AdS/QCD and completely out of
reach of chiral perturbation theory. For comparison, we also explore $f_2
\rightarrow \pi \pi$, which because of its sensitivity to the UV corrections
has much more uncertainty.  We also calculate the masses of the higher
spin resonances on the Regge trajectory of the $f_2$, and find they compare
favorably with experiment.
\end{abstract}

\end{titlepage}

\section{Introduction \label{SecIntro}}
It is well known that QCD, of the real world, cannot be studied through the
AdS/CFT correspondence. After all, QCD is not a conformal field theory, $N_c$
is not large, the string dual of QCD is a complete mystery, and if there
is such a dual, the string scale must be low. One can even make more practical
objections, such as that any low energy predictions which do come out cannot be original.
They must be the same as the predictions of chiral perturbation theory, since
the symmetries are the same. Or at least, since AdS/CFT employs the operator
product expansion, it should not be more powerful than QCD sum rules,
which should be able to extract as much information as possible out of the OPE. 
Therefore, why bother with AdS/QCD at all? We are
motivated by several reasons. First, with QCD we have experimental
results which can help determine the elements of AdS/CFT essential for 
establishing a predictive duality.   Second, after we
know what works and what does not, there is the possibility that we
may learn something about QCD itself, and perhaps about strongly coupled
gauge theories in more generality.  Finally, as our framework depends on
the holographic map, we are also indirectly exploring this map
experimentally. 

To elaborate on these motivations, we turn to the topic of the current work:
tensor mesons. We will focus on the $f_2$, a spin-two isospin singlet meson
with mass 1275 MeV, although higher spin mesons will be discussed as well. As
we will shortly see (Section \ref{SecCPT}), 
not much can be said about tensor mesons in
perturbation theory. Even if we couple the $f_2$ universally, assuming
general coordinate invariance (GC), there is an unknown dimensional coupling
constant, the analog of the Planck scale for the spin-2 graviton. The decay
rates to pions and to kaons can be related by $\mathrm{SU}( 3 )$, but the decay
rate to photons has a free NDA factor of order 1. The $f_2$ mass is  another free parameter.  
What's worse is that the cutoff on the
chiral Lagrangian is $\Lambda = 4 \pi f_{\pi} \approx 1200$ MeV, which is below
the $f_2$ mass. So even if chiral symmetry did make predictions, we would not
be able to trust them.

In contrast, on the AdS side, the $f_2$ will be treated as a spin-2 gauge field in the bulk
(Section~\ref{SecADS}). 
Equivalently, we can think of it as a KK excitation
of the graviton. Thus, we secure its interactions by GC. But now the coupling
constant can be calculated by matching to the perturbative OPE. The $f_2$ mass
is set by the eigenvalue equation for the KK mode. We define our units by
fixing the IR cutoff with the $\rho$ mass. This leads to the simple formula
that the ratio of the $f_2$ mass to the $\rho$ mass is simply given by the
ration of the zeros of 
Bessel functions $\mathcal{J}_1$ and $\mathcal{J}_0$. Thus
\begin{equation}
  \frac{m_{f_2}}{m_{\rho}} = \frac{\mathrm{zero\: of \:} \mathcal{J}_1
  }{\mathrm{zero\: of \:} \mathcal{J}_0} = 1.59 \quad ( \mathrm{AdS/QCD} )
\label{masspred}
\end{equation}
The experimental value for this ratio is 1.64, a difference of 3\% .
Remarkably, mass predictions can be extended up the Regge trajectory for
higher spin mesons, without a noticeable loss of accuracy 
(Section~\ref{SecHigher}).

The coupling to photons is also fixed in this theory -- it is determined by
the overlap of the flat photon wavefunction with that of the $f_2$. Thus we
can compute
\begin{equation}
  \Gamma ( f_2 \rightarrow \gamma \gamma ) = 2.71 \: \mathrm{keV} \quad (
  \mathrm{AdS} / \mathrm{QCD} ) \label{f2ggpred}
\end{equation}
The observed width is $ 2.60 \pm 0.24$ keV, so we get this rate correct to
within experimental error.  These two calculations
indicate that, for some reason, AdS/QCD is more accurate then we have any
right to expect. Because the decay to photons is sensitive to the admixture 
of the tensor glueball operator, we actually obtain some non-trivial information 
about the $f_2$ from the this calculation. This is discussed in
section~\ref{Secfgg}.

Not everything in the tensor sector works so well. We also calculate
the rate for
$f_2$ decaying to pions, still without introducing any new free parameters,
and find it too small by a factor of 4. This is more in line with our
naive expectations. It indicates that either some of our assumptions, such as
the those about boundary conditions, should be reconsidered, or that higher dimension
operators are relevant. In fact, we know roughly where the string scale is on
the AdS side. Some of the tensor modes, such as $a_2 ( 1320 )$, must be string
states, because they carry isospin. So we are at least very near the regime
where string corrections become important. Gravity corrections are also
important because we know the effective Planck scale for the $f_2$, and it is
also low. So it is not surprising that our calculations can receive stringy corrections. In
Section \ref{SecHO} we show that higher dimension bulk operators are very relevant to the
$f_2$ to pion decays. More importantly, we also show that higher dimension
bulk operators are \textit{not} relevant for the decay to photons, so we have
more reason to trust our $f_2 \rightarrow \gamma \gamma$ calculation.

\section{Introducing the $f_2$ \label{SecCPT}}
Before we describe the AdS/QCD construction, we review and elaborate on what
is known about the $f_2$ from other methods. In particular, we  discuss
chiral perturbation theory, which allows us to couple the $f_2$ to pions. 
We will also establish notation and present some formulae which will be used later on.

\subsection{Chiral Perturbation Theory}
In chiral perturbation theory, the pions are Goldstone bosons for a
spontaneously broken $\mathrm{SU} ( N_f )_L \times \mathrm{SU} ( N_f )_R$ global
symmetry~\cite{Pich:1995bw}. They are embedded in a matrix $\pi$ which is exponentiated to get
\begin{equation}
  U = e^{2 i \pi / f_{\pi}}
\end{equation}
This matrix $U$ transforms under $S U ( N_f )_L \times S U ( N_f )_R$
\begin{equation}
  U \rightarrow g_L U g_R^{\dag}
\end{equation}
and the non-linear transformations of the pions follow. The lowest order
chiral Lagrangian (with massless quarks) is
\begin{equation}
  \mathcal{L}^{( 2 )} = \frac{f_{\pi}^2}{4} \mathrm{Tr} [ D_{\mu} U^{\dag}
  D^{\mu} U ] = \frac{1}{2} ( \partial_{\mu} \pi^a ) ( \partial_{\mu} \pi^a )
  + \cdots
\end{equation}
To next order, there are additional terms~\cite{Gasser:1984gg}
\begin{equation}
  \mathcal{L}^{( 4 )} = L_1 \mathrm{Tr} [ D_{\mu} U^{\dag} D^{\mu} U ]^2 + L_2
  \mathrm{Tr} [ D_{\mu} U^{\dag} D_{\nu} U ] \mathrm{Tr} [ D^{\mu} U^{\dag}
  D^{\nu} U ] + L_3 \mathrm{Tr} [ D_{\mu} U^{\dag} D^{\mu} U D_{\nu} U^{\dag}
  D^{\nu} U ] \nonumber
\end{equation}
\begin{equation}
  + i L_9 \mathrm{Tr} [ F_R^{\mu \nu} D_{\mu} U D_{\nu} U^{\dag} + F_L^{\mu \nu}
  D_{\mu} U^{\dag} D_{\nu} U ] + L_{10} \mathrm{Tr} [ U^{\dag} F_R^{\mu \nu} U
  F^L_{\mu \nu} ]
\end{equation}
where $F_{R / L}^{\mu \nu}$ are the field strengths for external right- and
left-handed gauge fields. At this point, all the $L_i' s$ and $f_{\pi}$ are
unknown parameters to be fit to data. Often, assumptions additional to chiral
symmetry can predict some of these $L_i$'s. For example, vector meson
dominance assumes that the bare $L_i$ are all zero, and the observed $L_i$
come from integrating out heavy mesons, 
in particular the $\rho$ meson~\cite{Ecker:1988te,Ecker:1989yg,Harada:2003jx}. 
This predicts some relations between the $L_i$. Another example
 is AdS/QCD which
predicts similar, but different, 
relations~\cite{DaRold:2005zs,DaRold:2005vr,Hirn:2005nr}. 
The $L_i$ are fairly well known,
and are therefore handy for distinguishing different theories.

To add the $f_2$, we introduce, by hand, a massive spin-2 meson. Its kinetic
term should be of the Fierz-Pauli form~\cite{Fierz:1939ix}
\begin{equation}
  \mathcal{L}_{\mathrm{kin}}^{( f )} = \frac{1}{2} h_{\mu \nu} \Box h_{\mu \nu}
  +  h_{\mu \alpha, \alpha}^2 -  h_{\mu \nu, \mu} h_{,
  \mu} + \frac{1}{2} h_{, \mu}^2 + \frac{1}{2} m_f^2 ( h_{\mu \nu}^2 - h^2 )
\end{equation}
where $h = h_{\mu \mu}$. Then the most general set of interactions with the
chiral Lagrangian begins
\begin{equation}
  \mathcal{L}_{\mathrm{int}} = c_1 h_{\mu \nu} \mathrm{Tr} [ D_{\mu} U
  D_{\nu} U^{\dag} ] + c_2 h \mathrm{Tr} [ D_{\mu} U D_{\mu} U^{\dag} ] + c_3
  h_{\mu \nu} \mathrm{Tr} [ U^{\dag} F^R_{\mu \alpha} U F^L_{\alpha \nu} ] +
  \cdots \label{linth}
\end{equation}
By chiral symmetry alone, the $c_i$, like the $L_i$, are completely
undetermined.

To progress, we can assume that the $f_2$ couples universally to the energy
momentum tensor for matter, like in general relativity. And, like in general
relativity, this would let us bootstrap the $f_2$'s self-couplings as well, at
least in the absence of an $f_2$ mass. But because of the mass, the final
Lagrangian will have no exact symmetry lacking in the general expression~\eqref{linth}.
Nevertheless, the assumption that the $f_2$ couples to the energy momentum tensor of matter
is consistent within effective field theory, as it parametrically raises the
scale at which the $f_2$ interactions become strong~\cite{Arkani-Hamed:2002sp,Schwartz:2003vj}.

Adding an $f_2$ coupling strength $G_f$ (if $f_2$ were a graviton, $G_f \approx
M_P^{- 1}$) to the Lagrangian, we have
\begin{equation}
  \mathcal{L} = \mathcal{L}^{(2)} + \mathcal{L}_{\mathrm{kin}}^{( f )} +
  \frac{1}{2} G_f h_{\mu \nu} \Theta^{( 2 )}_{\mu \nu} + \cdots
\end{equation}
where $\Theta^{( 2 )}_{\mu \nu}$ is the energy momentum tensor following from
$\mathcal{L}^{( 2 )}$
\begin{equation}
  \Theta_{\mu \nu}^{( 2 )} = \frac{f_{\pi}^2}{4} \left[ \eta_{\mu \nu} (
  D_{\kappa} U ) ( D^{\kappa} U^{\dag} ) - 2 ( D_{\mu} U ) ( D_{\nu} U^{\dag}
  ) \right]
\end{equation}
This is of the form $\mathcal{L}^{( f )}_{\mathrm{int}}$ with $c_1 = - 2 c_2$. We
can integrate out the $f_2$ now, recalling that because of the Fierz-Pauli
mass its propagator is~\cite{vanDam:1970vg,VanNieuwenhuizen:1973fi}
\begin{equation}
  \langle h_{\mu \nu} h_{\rho \sigma} \rangle = \frac{P_{\mu \nu \rho
  \sigma}}{p^2 - m_f^2}
\end{equation}
where the projector, in the particle's rest frame, has the form
\begin{equation}
  P_{\mu \nu \rho \sigma} = \frac{1}{2} ( \eta_{\mu \rho} \eta_{\nu \sigma} +
  \eta_{\mu \sigma} \eta_{\nu \rho} - \frac{2}{3} \eta_{\mu \nu} \eta_{\rho
  \sigma} ) \label{proj}
\end{equation}
(In the massless case, or if the mass had just the $h_{\mu \nu}^2$ term, the
residue would have a $1$ instead of a $2 / 3$). Then integrating out the $f_2$
gives an $\mathcal{L}^{( 4 )}$ term with coefficients
\begin{equation}
  L_1^f = - \frac{G_f^2 f_{\pi}^4}{96 m_f^2} \quad \mathrm{and} \quad L_2 = -3 L_1
\end{equation}
Of course, these are not predictions, as the $L_i$ can have contributions from
integrating out other fields as well, or simply from adding a bare term.

We will find in the next paragraph, that from the $f_2$ decay rate into 
pions we can set $G_f = g_{f\pi\pi}= 0.019\: \mathrm{MeV}^{-1}$. 
This leads to, $L_1 = - 0.17 \times 10^{- 3}$ and $L_2 =
0.50 \times 10^{- 3}$. These are fairly substantial contributions to the
experimental values of $L_1 = 0.4 \pm 0.3 \times 10^{- 3}$ and $L_2 = 1.4 \pm
0.4 \times 10^{- 3}$. Of course, this tells us nothing about the $f_2$, as we
must make assumptions about what else contributes to the $L_i$, which are just
as likely to be wrong as our universality hypothesis for the $f_2$ couplings.

Now let us turn to this the $f_2$ decays. The $f_2$ decays predominantly into
pions. The minimal couplings can be written as
\begin{equation}
  \mathcal{L}_{f \pi \pi} = \frac{1}{2} g_{f \pi \pi} h_{\mu \nu}
  \Theta_{\mu \nu}^{\pi}, \quad   \Theta_{\mu \nu}^{\pi} = \frac{1}{2}
  \eta_{\mu \nu} ( \partial_{\alpha} \pi )^2 - ( \partial_{\mu} \pi ) (
  \partial_{\nu} \pi )
\end{equation}
then the decay rate is~\cite{Han:1998sg,Aliev:1981ju,Suzuki:1993zs}
\begin{equation}
  \Gamma ( f \rightarrow \pi^+ \pi^-, \pi^0 \pi^0 ) = g_{f \pi \pi}^2
  \frac{m_f^3}{1280 \pi} \left( 1 - 4 \frac{m_{\pi}^2}{m_f^2} \right)^{5 / 2}
  = g_{f \pi \pi}^2 ( 4.56 \times 10^5 \mathrm{MeV}^3 )
\end{equation}
where we have used $m_f = 1275.4$ MeV and $m_{\pi} = 139.5$ MeV. We are also
interested in the decay into photons, which is well measured also. If we
define the coupling as
\begin{equation}
  \mathcal{L}_{f \gamma \gamma} 
= \frac{1}{2} g_{f \gamma \gamma}  h_{\mu \nu} \Theta_{\mu \nu}^{\gamma}, 
\quad \quad 
\Theta_{\mu \nu}^{\gamma} = \frac{1}{4} \eta_{\mu \nu} F_{\alpha \beta}^2 - 
F_{\nu \alpha} F_{\alpha \mu} \label{gemt}
\end{equation}
then the formula for this decay rate is~\cite{Han:1998sg,Suzuki:1993zs}
\begin{equation}
  \Gamma ( f_2 \rightarrow \gamma \gamma ) = g_{f \gamma \gamma}^2
  \frac{m_f^3}{320 \pi} = g_{f \gamma \gamma}^2 ( 2.04 \times 10^6
  \mathrm{MeV}^3 )
\end{equation}
The observed decay rates are  $\Gamma ( f \rightarrow \pi \pi ) =$
156 MeV and $\Gamma ( f_2 \rightarrow \gamma \gamma ) =$ 2.6 keV which imply
$g_{f\pi \pi} = 0.0185$ MeV$^{- 1}$ and $g_{f \gamma \gamma} = 3.55 \times 10^{-
5}$ MeV$^{- 1}$. A helpful way to compare theory and experiment is through the
ratio of decay rates with the phase space factored out
\begin{equation}
  \frac{g_{f \pi \pi}^2}{g_{f \gamma \gamma}^2} = 2.71 \times 10^5 \quad (
  \mathrm{Exp} )
\end{equation}
Assuming strict universality of the $f_2$ couplings would lead to $g_{f \pi
\pi} = g_{f \gamma \gamma} = G_f$, and get this ratio wrong by six orders of
magnitude.

To get a better prediction, we must use the fact that the photon is really
external to QCD, and so its coupling must be suppressed by powers of the weak
coupling constant $e / 4 \pi$. Naive dimensional analysis~\cite{Manohar:1983md}
instructs us to take
$g_{f \gamma \gamma} = c_{\mathrm{NDA}} e^2 / 16 \pi^2 G_f$, for some constant
$c_{\mathrm{NDA}}$ of order one. Then
\begin{equation}
  \frac{g_{f \pi \pi}^2}{g_{f \gamma \gamma}^2} = \frac{1}{c_{\mathrm{NDA}}^2}
  \left( \frac{16 \pi^2}{e^2} \right)^2 = \frac{1}{c_{\mathrm{NDA}}^2} 2.96
  \times 10^6 \quad ( \mathrm{NDA} )
\end{equation}
This is a fairly good estimate. We see that $c_{\mathrm{NDA}} = 3.31$. But we
cannot get $c_{\mathrm{NDA}}$ by NDA or by assuming universality of the $f_2$
couplings (which would say $c_{\mathrm{NDA}} = 16 \pi^2 / e^2$).

One way to estimate $c_{\mathrm{NDA}}$ has been proposed in~\cite{Suzuki:1993zs}. 
Similar to the assumption in vector meson dominance, it declares that the decay to photons is
mediated purely by $\rho$ exchange. That is, $f$ decays into two $\rho$'s
which decay into photons. So the bare coupling $g_{f \gamma \gamma}$ is zero.
This assumption can be called tensor meson dominance 
(TMD)~\cite{Ishikawa:1987xi,Aliev:1981ju,Suzuki:1993zs}. 
Taking $g_{f \pi\pi}$ from experiment, it gives
\begin{equation}
  \frac{g_{f \pi \pi}^2}{g_{f \gamma \gamma}^2} = 3.6 \times 10^6 \quad (
  \mathrm{TMD} )
\end{equation}
This is not better than naive dimensional analysis with $c_{\mathrm{NDA}} = 1$.

Tensor meson dominance also relates the $f_2$ decay constant $\ff$ to $g_{f \pi\pi}$. 
The $f_2$ decay constant is defined by
\begin{equation}
\langle \Theta_{\mu\nu}| f_2\rangle = \ff m_f^3 \epsilon_{\mu\nu} \label{ffdef}
\end{equation}
where $\epsilon_{\mu\nu}$ is the polarization tensor for the $f_2$ and $\Theta_{\mu\nu}$ is
the energy momentum tensor. We have chosen $\ff$ to be dimensionless, and so it is off by
factors of $m_f$ from what is often called $g_f$ in the literature.
If we assume, via TMD, that the pion tensor form factors are
determined by $f_2$ exchange, even as $p \to 0$, then the normalization of the pion's energy determines
\begin{equation}
\ff = \frac{1}{m_f g_{f\pi\pi} } = 0.042 \quad (\mathrm{TMD}) \label{tmdff}
\end{equation}
We cannot compare $\ff$ to experiment, but we will compare this value to predictions from AdS/QCD 
below.

In summary, we wrote down the most general Lagrangian involving a spin-2 field,
the $f_2$, which respects chiral symmetry. We assumed the $f_2$ couples to
the energy momentum tensor of the various matter fields,
like the graviton, even though the $f_2$ is massive. Its mass, its
coupling constant, and an order one factor in its coupling to photons are all
free parameters. We can fit these to the observed mass, decay rate to pions,
and decay rate to photons, but then there are no predictions. Instead, we can
assume tensor meson dominance \textit{ad hoc}, which eliminates one parameter,
and makes a prediction which turns out not to much better than naive
dimensional analysis.

\section{AdS/QCD \label{SecADS}}
Now we turn to the predictions from AdS/QCD. We will mostly follow the
notation of \cite{DaRold:2005zs}, although \cite{Erlich:2005qh} 
presents an equivalent formulation. Our review
will be quick, and we refer the reader to these two papers, or to the original
AdS/CFT literature~\cite{Maldacena:1997re,Witten:1998qj,Gubser:1998bc,Aharony:1999ti}
for more details.

The basic premise of AdS/CFT is that there is a duality between a 4D conformal
field theory and a 5D gravity theory on a curved AdS background. Position in
the extra dimension corresponds to energy in the 4D theory. We will use
conformal coordinates, with the curvature scale normalized to 1, so the metric
is
\begin{equation}
  d s^2 = \frac{1}{z^2} ( d x^2_{\mu} - d z^2 )
\end{equation}
Energy independence in the CFT corresponds to an AdS isometry which shifts
$z$. Small $z$ is the high energy UV region, and large $z$ is the IR.  Though
approximately conformal in the UV (in the sense that 
$\beta \sim g^3 \to 0$),
QCD becomes strongly coupled, and breaks conformality in IR. We avoid 
this awkward region by explicitly
cutting off the space at $z = z_m$, where $z_m$ (which will be of order
$1/\Lambda_{QCD}$) is to be fit to data.  We thus implicitly assume that
asymptotic freedom sets in very quickly as we go the UV in QCD, motivating
our choice of the AdS metric.

\subsection{The $f_2$ mass \label{Secf2mass}}

Operators in the CFT, or in this case QCD, correspond to fields in the AdS
bulk. For example, the $\rho$, which is a isospin triplet spin-1 meson in QCD,
corresponds to a bulk gauge field $V_M$ in bulk. The action for $V_M$ is just
\begin{equation}
  \mathcal{L} = \int d^5 x \sqrt{g} \frac{1}{4} F_{M N} F^{\mathrm{MN}} = \int
  d^5 z \frac{1}{z} ( \frac{1}{4} F_{\mu \nu}^2 - \frac{1}{2} F_{\mu 5}^2 )
\end{equation}
More precisely, $\rho$ is the first KK excitation of the 4-vector part
$V_{\mu} ( x, z )$, while the other KK modes, $\rho_n$, correspond to heavier
resonances with the same quantum numbers. Thus the $\rho_n$ are 4D fields with
5D profiles determined by the vector KK equation in AdS
\begin{equation}
  z \partial_z \frac{1}{z} \partial_z V_n ( z ) + ( m^V_n )^2 V_n ( z ) = 0.
\label{veckk}
\end{equation}
The generic solution is
\begin{equation}
  V_n ( z ) = N_n z \left[ \mathcal{J}_1 ( m^V_n z ) + \beta_n \mathcal{Y}_1 (
  m^V_n z ) ] \right.
\end{equation}
We then impose boundary conditions $\rho' ( z_m ) =\rho(0)= 0$, 
which sets $\beta_n = 0$ and quantizes the masses to be solutions of
$\mathcal{J}_0 ( m_n^V z_m ) = 0$. The normalization is set by
 $\int d z \rho_n ( z )^2 / z = 1$ and so the $\rho$ wavefunction is
\begin{equation}
  V ( z ) = 2.72 \frac{z}{z_m} \mathcal{J}_1 ( 2.40 \frac{z}{z_m} )
\end{equation}
We can then use the observed $\rho$ mass of 775 MeV to fix $z_m = ( 323\: \mathrm{MeV}
)^{- 1}$.

For the $f_2$, we need a spin-2 field in the bulk, $h_{\mathrm{MN}}$. The first
KK mode of $h_{\mathrm{MN}}$ will be the $f_2$. Of course, we already have a
spin-2 field in the bulk, the graviton. Expanding linearized gravity excitations about
the AdS background
\begin{equation}
  d s^2 = \frac{1}{z^2} ( \eta_{\mu \nu} + h_{\mu \nu} ) d x^{\mu} d x^{\nu} -
  \frac{1}{z^2} d z^2 \label{lingrav}
\end{equation}
produces a Lagrangian
\begin{equation}
  \int d^5 x \sqrt{g} R_5 = \int d^5 x \frac{1}{2} \frac{1}{z^3} [ \partial_z
  h_{\mu \nu} \partial_z h_{\mu \nu} -  h_{\mu \nu} \Box h_{\mu \nu} + \cdots
  ] \label{haction}
\end{equation}
But since we are not demanding a fully consistent theory, we could just as
well have introduced a different spin-2 field. In fact, we expect from QCD
that there are a number of isospin singlet spin-2 particles, including
glueballs, and so a complete formulation should include a number of bulk
spin-2 fields.

Each field will have a tower of KK modes, whose profiles are determined by
the tensor equation of motion
\begin{equation}
  z^3 \partial_z \frac{1}{z^3} \partial_z h_n ( z ) + ( m_n^h )^2 h_n ( z ) =
  0. \label{tenkk}
\end{equation}
The generic solution is
\begin{equation}
  h_n ( z ) = N_n z^2 \left[ \mathcal{J}_2 ( m_n^h z ) + \beta_n \mathcal{Y}_2
  ( m_n^h z ) ] \right.
\end{equation}
Boundary conditions $h'(z_m)=h(0)=0$ again put $\beta_n = 0$ and now quantize the
masses according to $\mathcal{J}_1 ( m_n^h z_m ) = 0$. For the spin-2 case,
the normalization is set by $\int d z h_n^2 / z^3 = 1$. Thus the $f_2$ profile
is
\begin{equation}
  h ( z ) = 3.51 \frac{z^2}{z_m} \mathcal{J}_2 ( 3.83 \frac{z}{z_m} ) \label{hwave}
\end{equation}
Thus we get our first prediction. The $f_2$ mass is predicted to be 
3.83 $z_m^{- 1} =$ 1236 MeV, which, as noted in the introduction is, only 
$ 3\% $ off of the observed mass of 1275 MeV.  In other words, we
have made the $z_m$ independent prediction $m_{f_2}/m_{\rho} = 1.59$ of
equation~\eqref{masspred}. The experimental central values give
\begin{equation}
\frac{m_{f_2}}{m_\rho} = 1.64 \quad (\mathrm{Exp})
\end{equation}

To be fair, we have been working
in the free field approximation, so that, at this point, the $f_2$ and $\rho$
are infinitely narrow resonances. In fact, the widths of the $f_2$ and $\rho$ are
$185$ MeV and $146$ MeV respectively, and thus there is an uncertainty in what
we should expect the resonance mass to be. 
So the AdS/CFT prediction is entirely satisfactory.
It is remarkable how little we had
to introduce to arrive at this prediction -- no chiral symmetry breaking, no mention
of the number of flavors or colors, in fact, no matching to QCD at all.
We simply assumed that the lightest spin-2 mode
is captured by a 5D massless spin-2 field (i.e. the dual of a spin-2 operator
of the lowest dimension).

\subsection{Matching to QCD \label{SecMatch}}
Next, we will calculate the coupling constant of the $f_2$ by matching to the
operator product expansion for the tensor-tensor two point function in QCD. We
prefer to express all the dimensionful scales in terms of $z_m$, so we write
the coupling constant as $G_f = z_m g_f$, leaving $g_f$ as the dimensionless
coupling constant associated with the $f_2$, directly analogous to $g_5$ for
the $\rho$.

For each quark, we can construct a tensor bilinear,
\begin{equation}
  \Theta^{\mu \nu}_q 
= \frac{1}{4} i \bar{q} ( \gamma^{\mu}  \overset{\leftrightarrow}{\partial}_\nu + \gamma^{\nu} \overset{\leftrightarrow}{\partial}_\mu
  ) q
\end{equation}
where $\overset{\leftrightarrow}{\partial} = \overset{\leftarrow}{\partial} -
\overset{\rightarrow}{\partial}$. This is just the energy momentum tensor for quarks in
QCD~\cite{Han:1998sg}. The OPE we need is the $\Theta \Theta$ two-point function, which can be
written as
\begin{equation}
  \langle \Theta^{\mu \nu}_q \Theta_q^{\rho \sigma} \rangle = P^{\mu \nu \rho
  \sigma} \Pi_{\theta \theta}^q ( p^2 )
\end{equation}
where $P^{\mu \nu \rho \sigma}$ is the transverse projector \eqref{proj}. The
constant term in the $\Theta \Theta$ OPE comes from a 1-loop quark 
diagram~\cite{Aliev:1981ju,Reinders:1981ww}
\begin{equation}
  \Pi_{\theta \theta}^q ( p^2 ) 
= - \frac{N_c}{160 \pi^2} p^4 \ln (\frac{p^2}{\Lambda^2} ) + \cdots \label{qcdtt}
\end{equation}

Since we have the correlator as a function of momentum $p$ for fixed cutoff
$\Lambda$, on the 5D side, we should think of the current as a source on the
UV boundary at $z = \varepsilon$,
\begin{equation}
  \mathcal{L} \supset \frac{1}{2} g_f \delta ( z - \varepsilon ) h_{\mu \nu}
  \Theta^{\mu \nu}
\end{equation}
Then the  $\Theta \Theta$ correlator can be derived from an effective 4D
action at $z = \varepsilon$, with the fifth dimension integrated out:
\begin{equation}
  \langle \Theta^{\mu \nu} \Theta^{\rho \sigma} \rangle = \frac{4}{g_f^2}
\left.  \left( \frac{\delta}{\delta h_{\mu \nu}} \frac{\delta}{\delta h_{\rho
  \sigma}} e^{i S_{\mathrm{eff}}}  \right) \right|_{h = 0}
\end{equation}
The easiest way to compute this is through the bulk-to-boundary propagators $K(z)$, 
which propagates a source via $h_{\mu \nu}(z)=h_{\mu\nu}(\varepsilon)K(z)$. Then the action
\eqref{haction} reduces to a boundary term on the equations of motion, and we have simply
\begin{equation}
\langle \Theta^{\mu \nu} \Theta^{\rho \sigma} \rangle = P^{\mu \nu \rho\sigma} 
\frac{4}{g_f^2} \frac{K'(\varepsilon)}{\epsilon^3} \label{ttkp}
\end{equation}
For the purpose of evaluating $g_f$, the IR boundary conditions are irrelevant and
taking $z_m \to \infty$ we solve for $K(z)$ directly. The solution involves the second order Bessel
function ${\mathcal{K}}_2$~\cite{Randall:2001gb}
\begin{equation}
K(p,z) =   \frac{z^2\mathcal{K}_2 ( p z  )}{\varepsilon^2 \mathcal{K}_2 (p \varepsilon)}
\end{equation}
It is not hard to check that this satisfies 
$( z^3\partial_z \frac{1}{z^3} \partial_z + p^2 ) K ( p, z ) = 0$ and $K(\varepsilon)=1$.
Thus the AdS prediction for the leading log in the tensor-tensor correlator is
\begin{equation}
 \Pi_{\theta \theta} ( p^2 ) = \frac{4}{g_f^2} \frac{K'(\varepsilon)}{\varepsilon^3}
= - \frac{1}{2 g_f^2} p^4 \log p^2  + \cdots
 \label{adstt}
\end{equation}
The right hand side of this equation presents the leading log 
in the small $z$ expansion of $\mathcal{K}_2(z)$.
Once we identify the quark content of the $f_2$, we can match the logs in \eqref{adstt} and \eqref{qcdtt} 
to fix $g_f$.

To identify the $f_2$ note that there is more than one isospin singlet state
we can construct in QCD. First, there are the quark states. But in addition,
in QCD there is a tensor glueball state. Its energy momentum tensor is
\begin{equation}
  \Theta_{\mu \nu}^G = \frac{1}{4} \eta_{\mu \nu} ( G_{\alpha \beta}^a )^2 -
  G_{\nu \alpha}^a G_{\alpha \mu}^a
\end{equation}
Conveniently, the two-point function for the glueball is QCD has also been
studied. Its OPE begins~\cite{Novikov:1981xj}
\begin{equation}
  \langle \Theta_{\mu \nu}^G \Theta_{\rho \sigma}^G \rangle = - P^{\mu \nu
  \rho \sigma} \frac{1}{10 \pi^2} p^4 \ln p^2 + \cdots 
\end{equation}
Now, in general, there will be kinetic mixing among the glueball and the quark
states, proportional to $\alpha_s$. Although, the explicit mixing is known, we
will be content to observe that one of the eigenstates should couple to the
conserved current, the energy momentum tensor. This will be the lightest
state, as the $\alpha_s$ corrections can only lift the mass, so we identify it
as the $f_2$.

Thus we are led to take the $f_2$ coupling to be
\begin{equation}
  \mathcal{L} = \frac{1}{2} g_f h_{\mu \nu} \Theta^{\mu \nu}, \quad
  \Theta_{\mu \nu} = 
\Theta_{\mu \nu}^u + \Theta_{\mu \nu}^d  + \Theta_{\mu\nu}^s + \Theta_{\mu \nu}^G 
\end{equation}
Hence, from QCD,
\begin{equation}
  \langle \Theta_{\mu \nu} \Theta_{\rho \sigma} \rangle = - P^{\mu \nu \rho
  \sigma} (\frac{N_c N_f}{160 \pi^2} +  \frac{N_c^2-1}{80 \pi^2}) p^4 \ln p^2 
\label{gluett}
\end{equation}
Taking $N_c=N_f=3$ and matching \eqref{adstt} produces
to \eqref{gluett} 
\begin{equation}
  g_f = \frac{4 \pi}{\sqrt{5}} .
\end{equation}
And using  $z_m^{- 1} = 322$ MeV, determined from the $\rho$ mass, the $f_2$
coupling constant is
\begin{equation}
  G_f = g_f z_m = 0.0174\: \mathrm{MeV}^{- 1}
\end{equation}
This coupling cannot be directly measured, but it lets us calculate the $f_2$ decay rates.

It is instructive to calculate the $f_2$ decay constant $\ff$ as well, defined in \eqref{ffdef}.
$\ff$ also appears in residue of the decomposition of
the two point function $\Theta \Theta$ into meson resonances. 
\begin{equation}
\langle \Theta \Theta \rangle = \sum_\mathrm{mesons} \frac{f_n^2 m_f^4}{p^2-m_n^2}
\end{equation}
On the AdS side, the easiest way to compute the residue is by observing that the 
Dirichlet Green's function has a similar decomposition over KK modes
\begin{equation}
G(z,z') =  \sum_\mathrm{KK\: modes} \frac{h_n(z) h_n(z')}{p^2-m_n^2}
\end{equation}
This Green's function is related to the bulk-to-boundary propagator by
$K(z')=(z_m^2/z^3)\partial_{z} G(z,z')|_{z=\varepsilon}$ 
and so using \eqref{ttkp} we can relate
the decay constant directly to the wavefunction.
For the $f_2$, with wavefunction~\eqref{hwave}, we get
\begin{equation}
\ff = \frac{2}{g_f}\frac{1}{m_f^3 z_m^3} \frac{h'(\varepsilon)}{\varepsilon^3}
=  \frac{8.47}{m_f^3 z_m^3 g_f} = 0.024
\end{equation}
This also cannot be measured, but we can compare it to the prediction from TMD 
($\ff = 0.042$). We will discuss the significance of this discrepancy after calculating the decay rates.

\subsection{$f_2 \rightarrow \gamma \gamma$ \label{Secfgg}}
We now discuss how to introduce the photon and calculate $\Gamma (f_2 \rightarrow \gamma \gamma)$. 
The photon is a vector
field which is external to QCD. Nevertheless, we can consider it as a bulk
gauge field, whose zero mode, the photon, is massless. Thus the photon will
have a flat profile in the bulk
\begin{equation}
  \gamma ( z ) = c_{\gamma}
\end{equation}
Generally, a flat mode in AdS is not normalizable. In fact, regulating with a
UV cutoff at $z = \varepsilon$
\begin{equation}
  \int_{\varepsilon}^{z_m} \frac{1}{z} \gamma ( z )^2 d z = c_{\gamma}^2 \log
  \frac{z_m}{\varepsilon} .
\end{equation}
Normally, this would be a serious problem, as this divergent normalization
appears as the coefficient of the photon kinetic term in the 4D action when
the extra dimension is integrated out
\begin{equation}
  \mathcal{L} = c_{\gamma}^2 \log \frac{z_m}{\varepsilon} \left( \frac{1}{4}
  F_{\mu \nu}^2 \right) + \cdots
\end{equation}
However, the coefficient of the term is supposed to be $1 / e^2$, and the
electric charge is divergent -- it has a Landau pole. Explicitly, including
only the quark contributions
\begin{equation}
  \frac{1}{e^2 ( \mu )} = \left( ( \frac{2}{3} )^2 + ( \frac{1}{3} )^2 + (
  \frac{1}{3} )^2 + \cdots \right) \frac{N_c}{6 \pi^2} \ln (
  \frac{\mu}{\Lambda} )
\end{equation}
But this means that if we set $c_{\gamma}^2 = e^2 / ( 3 \pi^2 )$ (for $N_f =
3$) then the photon kinetic term becomes canonically normalized, $-
\frac{1}{4} F_{\mu \nu}^2$. We can use this divergent $c_{\gamma}$ in
calculations by just taking the renormalized value for the electric charge at
the relevant energy scale. 
It is satisfying that in contrast to chiral
perturbation theory, where the electric charge must be added by hand,
including an order one uncertainty, here the electric charge appears naturally
from matching to QED.

Another way to model the photon is as a linear combination of the diagonal generators
of  $SU(3)_\mathrm{flavor}$. Its coupling can then be fit from the AdS/CFT matching 
to the two-point function of vector currents in QCD. Since the relevant diagram is
the same as the one-loop contribution as in the QED $\beta$-function, 
it leads to the same value for $c_\gamma$. Although the two ways of modeling the photon 
are equivalent, they have complimentary advantages. 
The former nicely demonstrates that the $z$-dependence of the photon
profile corresponds to its running coupling, while the latter shows that it is really only
the quark contributions which we must include. For example, we do not include the 
electron contribution to $c_\gamma$ because the electron is not part of our AdS/QCD model.

The coupling of the $f_2$ to the photon is determined by general coordinate
invariance. Using the parametrization \eqref{lingrav}, we get
\begin{equation}
  \mathcal{L} = \int d^5 x \sqrt{g} F_{M N} F^{\mathrm{MN}} = \frac{1}{2} g_{f
  \gamma \gamma} h_{\mu \nu} \Theta_{\mu \nu}^{\gamma} + \cdots
\end{equation}
where  $\Theta_{\mu \nu}^{\gamma}$ is given in \eqref{gemt} 
and the coupling constant
is determined by the overlap integral
\begin{equation}
  g_{f \gamma \gamma} = g_f \int_0^{z_m} \frac{d z}{z} h ( z ) \gamma ( z )^2
  = 0.671 c_{\gamma}^2 z_m g_f = 0.671 \frac{e^2}{3 \pi^2} z_m g_f =
  2.08 \times 10^{- 3} G_f
\end{equation}
If we use the value of $G_f$, calculated in the previous subsection by
matching to QCD (including the 3 quark and gluon contributions), this leads
to $g_{f \gamma \gamma} = 3.62 \times 10^{- 5}$ MeV$^{- 1}$ and a decay rate
of
\begin{equation}
  \Gamma ( f_2 \rightarrow \gamma \gamma ) = 2.70\: \mathrm{keV} \quad (
  \mathrm{AdS} / \mathrm{QCD} )
\end{equation}
Compared to the experimental rate of $2.60 \pm 0.24$ this is quite remarkable.
All we needed to get this rate was $N_c$ and $N_f$, and the $\rho$ mass to
define MeV. In fact, we do not even need the $\rho$ at all. If we take $z_m$
from the measured $f_2$ mass, $z_m = 333$ MeV, then $\Gamma ( f_2 \rightarrow
\gamma \gamma ) = 2.54$ keV, which is also within experimental error. 

At this point, since we have achieved a very favorable comparison to experiment, it
is useful to return to some of our assumptions. Recall that we 
we identified the $f_2$ with the energy-momentum tensor, because it is
lightest spin-2 field in the theory. Thus, in our model, the $f_2$ is part
glueball, and equal parts up, down, and strange. If fact, some quark models suggest that
the $f_2$ and its nonet partner, the $f_2'(1525)$, are close to ideally mixed. That is,
the $f_2$ is mostly up and down, and the $f_2'$ is almost pure strange. 
We have not attempted to include quark masses at this point, so we cannot split the $f_2$ from
the $f_2'$, but we can nevertheless compare our result to the results of making other
assumptions about the $f_2$ quark content. For example, if we decouple the strange, so
the $f_2$ is up, down and glue, we would get $\Gamma(f_2 \rightarrow \gamma\gamma)=2.13$ keV. It
is reassuring that the 3 quark value ($2.70$~keV) and the 2 quark value ($2.13$~keV) straddle
the experimental decay rate~($2.60 \pm 0.24$~keV). If we had not included the glueball component, by
ignoring the last term in Eq. \eqref{gluett}, the decay rates would be $7.51$~keV and $7.82$~keV for the 3 quark
and 2 quark cases respectively. This lets us conclude that the $f_2$ must have a large glueball component, consistent
with a universal coupling to the energy momentum tensor.

\subsection{$f_2 \rightarrow \pi \pi$}
Thus far, we have derived two predictions -- the $f_2$ mass and
the partial width $\Gamma(f_2 \rightarrow \gamma \gamma)$ -- and the only input from
experiment (besides $N_c$ and $N_f$) was the $\rho$ mass, which just set the
scale. To work out the rate $f_2 \rightarrow \pi \pi$ we need more
information. We need to model chiral symmetry breaking, including both the
quark condensate and quark masses. Needless to say, the construction we will
use is not unique, and ambiguities make these predictions less accurate, and
less convincing. Since this is not the main focus of this paper, we will
quickly review, and then simply use, the formalism suggested 
in~\cite{DaRold:2005zs}.

The global symmetries $\mathrm{SU} ( 3 )_L \times \mathrm{SU} ( 3 )_R$ of QCD map
to gauge symmetries in the AdS. So we introduce bulk gauge bosons $A^L_M$ and
$A^R_M$ for the two subgroups. In 4D, pions appear as the Goldstone bosons
when the global chiral symmetry is spontaneously broken. Since in 5D the
global symmetry is gauged, the Goldstone bosons are eaten and show up in
unitary gauge as a mass term $m_A^2 A_M^2$ for the axial vector boson $A_M =
A^R_M - A_M^L$. In the 4D description of the 5D Higgs mechanism, the vectors
$A_{\mu}$ eat the scalar $A_5$ components. So in order to have massless 4D
pions, we must arrange for $A_5$ to have a massless excitation.

The mass $m_A$ corresponds to the formation of vacuum condensates, such as
$\langle \bar{q} q \rangle$ in QCD. Thus we should think of $m_A$ as being
spontaneously generated by the expectation value of some bulk scalar field
$X$. Since this breaking occurs in the IR, the mass should be localized near
the IR brane. However, it should have some $z$ dependence corresponding to the
energy dependence of the operator $\langle \bar{q} q \rangle$. AdS/CFT tells
us to match the dimension of the 4D operator to the bulk mass of a 5D field.
This is somewhat intuitive -- the energy dependence of a operator is
determined by its scaling dimension, and the $z$ dependence of a bulk field by
its mass. So we assume $\langle \bar{q} q \rangle$ is represented by the
expectation value of a bulk scalar field $X$.
The scalar equations of motion in AdS lead to
\begin{equation}
  \langle X \rangle = m z + \xi z^3 \equiv v ( z )
\end{equation}
Spontaneous symmetry breaking occurs in the IR, at large $z$, so $\xi$
represents the strength of this effect. Similarly, 
explicit symmetry breaking due to quark masses, which is apparent even in the UV,
corresponds to the $m$ term, which is more relevant at small $z$. We will work in the
massless quark limit for simplicity, and thus set $m = 0$. Then the free
parameter $\xi$ corresponds to the chiral symmetry breaking scale
$\Lambda_{\mathrm{QCD}}^3 \sim \text{$\langle \bar{q} q \rangle$}$.

So, the bulk action for the gauge fields in the AdS background is
\begin{equation}
  \mathcal{S} = \int d^4 x d z - \frac{1}{4 z} ( F_{\mathrm{MN}}^L )^2 -
  \frac{1}{4 z} ( F_{\mathrm{MN}}^R )^2 + \frac{v ( z )^2}{2 z^3} ( A_M )^2
\end{equation}
Note that the vector combination $V_M = A_M^L + A_M^R$, does not have a bulk
mass. So its first KK mode will be the same $\rho$ we introduced previously.
For the axial section, we want to decouple $A_5$ and $A_{\mu}$. To do this, we
introduce Goldstone bosons and gauge-fix, following~\cite{DaRold:2005zs}.
This separates the pions from the axial vectors.

After this gauge fixing, the KK equation for the axial vector is
\begin{equation}
  \left[ m^2 - 2 \frac{v ( z )^2}{z^2} + z \partial_z \frac{1}{z} \partial_z
  \right] a_n ( z ) = 0
\end{equation}
We can use this and the experimental $a_1$ mass to fix $\xi$. With boundary
conditions $a' ( z_m ) = a' ( 0 ) = 0$, and using $z_m^{- 1} = 322 \mathrm{MeV}$
from the $\rho$ mass, we find, numerically, an eigenvalue at the $a_1$ mass of
1230 MeV for $\xi = 3.77 z_m^{- 3}$.

The zero mode of $A_5$ represents the physical pions. The KK profile for this
mode satisfies
\begin{equation}
  \partial_z \left( \frac{z^3}{2 v ( z )^2} \partial_z \left( \frac{1}{z} A_5
  ( z ) \right) \right) = A_5 ( z )
\end{equation}
The solution is
\begin{equation}
  A_5 ( z ) = N \xi z^3 \left[ \mathcal{I}_{2 / 3} ( \frac{\sqrt{2}}{3} \xi
  z^3 ) - \beta_{\pi} \mathcal{K}_{2 / 3} ( \frac{\sqrt{2}}{3} \xi z^3 )
  \right]
\end{equation}
Although we have not imposed boundary conditions, $A_5 ( 0 ) = 0$
automatically. This is essentially set by the fact that we insist on having a
massless mode. The boundary condition $A_5 ( z_m ) = 0$ then sets
$\beta_{\pi}$. The normalization is determined by a $z$ integral
\begin{equation}
  - \int d z \left[ \frac{1}{z} A_5 ( z )^2 + \frac{1}{z} \frac{z^3}{2 v ( z
  )^2} ( \partial_z \frac{1}{z} A_5 ( z ) )^2 \right] = 1
\end{equation}
On the solution for $A_5 ( z )$, the expression in brackets is a total
derivative, which lets us calculate the normalization analytically
\begin{equation}
  N^{- 1 / 2} = \frac{\mathcal{I}_{2 / 3} ( \frac{\sqrt{2}}{3} \xi z^3_m )}{2
  \mathcal{K}_{2 / 3} ( \frac{\sqrt{2}}{3} \xi z^3_m )^2} \left[ \sqrt{3} \pi
  \mathcal{I}_{2 / 3} ( \frac{\sqrt{2}}{3} \xi z^3_m ) + 3 \mathcal{K}_{2 / 3}
  ( \frac{\sqrt{2}}{3} \xi z_m^3 ) \right]
\end{equation}
Thus, identifying this mode with the pion wavefunction, we have
\begin{equation}
  \pi ( z ) = N \xi z^3 \left[ \mathcal{I}_{2 / 3} ( \frac{\sqrt{2}}{3} \xi
  z^3 ) - \frac{\mathcal{I}_{2 / 3} ( \frac{\sqrt{2}}{3} \xi z^3_m
  )}{\mathcal{K}_{2 / 3} ( \frac{\sqrt{2}}{3} \xi z^3_m )} \mathcal{K}_{2 / 3}
  ( \frac{\sqrt{2}}{3} \xi z^3 ) \right]
\end{equation}
Now we can calculate $g_{f \pi \pi}$.

The coupling of $f_2$ to the pions is determined by general coordinate
invariance. Since everything is canonically normalized, we simply evaluate the
overlap of the $f_2$ with the combination that appears in the kinetic term for
$\pi$.
\begin{equation}
  g_{f \pi \pi} = g_f \int d z h ( z ) \left[ \frac{1}{z} \pi ( z )^2 +
  \frac{z^3}{2 v ( z )^2} ( \partial_z \frac{1}{z} \pi ( z ) )^2 \right] =
  0.519 z_m g_f = 0.519 G_f \label{gfpp}
\end{equation}
So we predict for the $G_f$-independent ratio
\begin{equation}
  \frac{g_{f \pi \pi}^2}{g_{f \gamma \gamma}^2} = 6.24 \times 10^4 \quad (
  \mathrm{AdS} / \mathrm{QCD} )
\end{equation}
Compared to the experimental value of $2.72 \times 10^5$ this is off by a
factor of 4. Using our previous result, $G_f = 0.0174$ MeV${}^{-1}$, 
we get
$g_{f \pi \pi} = 9.05 \times 10^{- 3}$ MeV${}^{- 1}$, and predict
\begin{equation}
  \Gamma ( f_2 \rightarrow \pi^+ \pi^-, \pi^0 \pi^0 ) = 37.4 \; \mathrm{MeV} \quad
  ( \mathrm{AdS} / \mathrm{QCD} )
\end{equation}
This also off  from the experimental value of $156.9 \pm 3.8$ MeV by a factor
of 4.

This factor of 4 is from the same origin as the factor of $2$ discrepancy between $f_f$ from AdS/QCD 
($f_f\sim 0.02$) and $f_f$ from TMD. After all, in TMD, $f_f$ is 
extracted from the $f_2 \to \pi\pi$ rate. 
In AdS/QCD, there is a conflict between the decay rates to photons and to pions, because
they depend on the same parameters;  but we will now see that it is justifiable, and it is
good that the photon rate is the one which comes out right.

\section{Higher order operators \label{SecHO}}
We can understand why the decay to photons was so accurate, but the decay to
pions was so far off by looking at higher dimension operators. Because the
$f_2$ is heavy, $1275$ MeV, compared to the naive cutoff of the chiral
Lagrangian, $4 \pi f_{\pi} = 1168$ MeV, we expect chiral perturbation theory
not to be accurate in the $f_2$ sector.
We can also see the breakdown of perturbation theory on the AdS side, by
looking at where higher dimension derivative operators become relevant.
Let us now
consider some of these operators, relevant to the $f_2 \rightarrow \pi \pi$
process.

Since we are interested in $f_2 \rightarrow \pi \pi$ we want operators linear in
$h$ and quadratic in $A_M$. The term we have been using to calculate $g_{f \pi
\pi}$ was determined by general coordinate invariance and the $\pi$ kinetic
term. It is, roughly,
\begin{equation}
  \int d^5 x \frac{1}{z} g_f h_{\mu \nu} ( \partial_{\mu} A_5 ) (
  \partial_{\nu} A_5 ) = ( 0.209 ) G_f \int d^4 x h_{\mu \nu} (
  \partial_{\mu} A_5 ) ( \partial_{\nu} A_5 ) \label{ho0}
\end{equation}
The number 0.209 comes from the $z$-integral, with the $f_2$ and $\pi$
wavefunctions. The difference from \eqref{gfpp} 
is because we have simplified the
tensor structure, but it is the same order of magnitude.

A possible higher order term, leading to exactly the same 4D structure is
\begin{equation}
  \int d^5 x \frac{1}{z} g_f h_{\mu \nu} ( \partial_{\mu} A_5 ) ( z^2
  \partial_z^2 ) ( \partial_{\nu} A_5 ) = ( - 1.57 ) G_f \int d^4 x h_{\mu
  \nu} ( \partial_{\mu} A_5 ) ( \partial_{\nu} A_5 ) \label{ho1}
\end{equation}
This could come from the 5D general coordinate invariant term
\begin{equation}
  \int d^5 x \sqrt{g} F_{\mathrm{MN}} D^N D^P F_{\mathrm{PM}}
\end{equation}
Or
\begin{equation}
  \int d^5 x \frac{1}{z} g_f ( z^2 \partial_z^2 h_{\mu \nu} ) (
  \partial_{\mu} A_5 ) ( z^2 \partial_z^2 ) ( \partial_{\nu} A_5 ) = (
  5.05 ) G_f \int d^4 x h_{\mu \nu} ( \partial_{\mu} A_5 ) (
  \partial_{\nu} A_5 )
\end{equation}
which might come from
\begin{equation}
  \int d^5 x \sqrt{g} R_{\mathrm{MNOP}}(h) F^{MN} F^{OP},
\end{equation}
where $R_{\mathrm{MNOP}}(h)$ is invariant under $h_{\mu\nu} \rightarrow
h_{\mu\nu} + \partial_\mu \xi_\nu(x) + \partial_\nu \xi_\mu(x)$.  Note
that linearized $\tilde{R}_{\mu\nu}(h)=R_{\mu\nu}(h)-4/z^2 h_{\mu\nu}$ 
(the combination invariant under the above shift) vanishes when the $f_2$ is on shell.

We could assume that these terms are suppressed compared to the leading piece
by some factors of the weak coupling constant $g_f$. At best, we could use the
5D loop factor, and suppress by $g_f^2 / 24 \pi^3 = 0.04$. Thus the 
correction~\eqref{ho1} 
contributes $30\%$ relative to the original term~\eqref{ho0}, 
which is not a small effect. With the 4D loop
factor, $16 \pi^2$, the new term is 1.5 times as important as the original.
Either way, there is no excuse to ignore the higher order corrections and we
cannot reasonably expect our rate $f_2 \rightarrow \pi \pi$ to be at all
accurate. In fact, it is surprising that we were even able to get the coupling
constant to a factor of 2.

In contrast, consider the terms contributing to corrections to $f_2
\rightarrow \gamma \gamma$. At leading order, we had
\begin{equation}
  \int d^5 x \frac{1}{z} g_f h_{\mu \nu} F_{\mu \alpha} F_{\nu \alpha} 
= (0.671 ) G_f \int d^4 x h_{\mu \nu} F_{\mu \alpha} F_{\nu \alpha} \label{fgg671}
\end{equation}
Any correction to the $f_2 \rightarrow \gamma \gamma$ rate has to come from a
vertex with one $h$ and two $F' s$. But the photon profile is flat, so any
term with $\partial_z$'s acting on $F$ will vanish. Terms with
$\partial_{\mu}$ acting on $F$ (like the Riemann term above)
may appear, but they cannot contribute to the
decay rate because the photon is massless and transverse -- there is nothing
to which we may contract the new momentum factor. 

One might worry that term such as
\begin{equation}
\int d^5 x \sqrt{g} F_{\mathrm{MN}} D^2 F^{\mathrm{MN}} \label{ddff}
\end{equation}
could contribute  if the derivatives are $\partial_z$'s 
and they act on the background metric.
But the effect is then only to shift the coupling constant $g_5$, which is unobservable
because the renormalized coupling has been matched to QCD. Another way
to see this is that 
on shell $D^2 F_{\mathrm{MN}} \sim 1/z^2 F_{\mathrm{MN}}$ which is the standard 
gauge kinetic term.
Thus, the contribution of~\eqref{ddff} has already been accounted for 
in the definition of $g_5^2$.
So no operators, analogous
to the ones relevant for the pion decay, affect the photon decay mode at all.

Throughout, we have been ignoring boundary operators on the IR brane. But these
can affect the photon decay rate, and should be included in a consistent effective field
theory. The natural size of a term like $\delta(z-z_m)h F^2$ is of order 
$z_m$ times $g_f^2 /16 \pi^2 = 0.20$, or about $.04 G_f$. 
Comparing to~\eqref{fgg671}  we expect corrections of order $10\%$ to the $f_2$ 
photon coupling. In addition, some other assumption could be wrong. For example,
$\alpha_s$ corrections have not been included -- we have assumed
that QCD is conformal. So this can change the result as well. 
Really, we could not have known ahead of time that we would get
$f_2 \rightarrow \gamma \gamma$ so precisely. Nevertheless, the rate came
out right, and it is suggestive that the contribution of boundary operators
and violations of conformality are small.

\section{Higher Spin States \label{SecHigher}}
In addition to the $f_2$, which we have studied here, QCD has a whole tower of
high spin resonances (for example, the $f_6$ with spin 6 sits at 2510 MeV).
In this section, we derive the predictions of AdS/QCD for their masses, and compare
to experiment.

A spin $s$ state in QCD corresponds to a field $\phi_{M_1 M_2...M_s}$ in AdS.
Ideally, we would like to write down an action for $\phi$, but higher spin Lagrangians are
in general very complicated.
For example, even at the free field level in flat space,
one needs to introduce on the order of $s$ auxiliary fields to eliminate the 
propagation of unphysical modes. Instead, we will take a more direct approach,
and simply use symmetry arguments to guess the equations of motion for 
the transverse spin-$s$  modes.

It is natural to expect the AdS theory for $\phi$ to respect a 
linearized gauge invariance, under which
\begin{equation}
\phi_{M_1 M_2...M_s} \rightarrow \phi_{M_1 M_2 \ldots M_s} 
+ D_{\{M_1} \xi_{M_2 \ldots M_s\}},
\end{equation}
Such a symmetry allows us to go to an axial gauge in which $\phi_{5 M_2 \ldots M_s}=0$.
In AdS, it is not hard to show that this gauge is 
preserved under residual transformations with $\xi_{5 M_3\ldots M_s}=0$
and $\xi_{\mu_2 \ldots \mu_s}(x,z) = z^{2-2s}\xi_{\mu_2 \ldots \mu_s}(x)$. This means that
a field $\phi(x,z) = z^{2-2s} \tilde\phi (x)$ 
simply shifts under the 4D gauge transformation, and therefore represents a zero mode. 
In terms of $\tilde \phi$, in this axial gauge, the Lagrangian contains a piece 
$\sqrt{g} (\partial_\mu\phi)^2 = z^{1-2s} (\partial_\mu\tilde\phi)^2$ and a similar
piece with $\partial_z$'s acting directly on $\tilde \phi$.
Thus the KK modes for a field of spin $s$ will satisfy
\begin{equation}
\partial_z \frac{1}{z^{2s-1}} \partial_z \tilde{\phi} +\frac{m^2}{z^{2s-1}} \tilde{\phi} =0 
\label{spinkk}
\end{equation}
Note that this equation is just a naive generalization of \eqref{veckk} and \eqref{tenkk} and matches
previous results for spins $1$ and $2$.

We can confirm this result from AdS/CFT reasoning as well. 
On the QCD side, the lowest dimension operator of spin $s$,
may contain terms like
\begin{eqnarray}
&& \bar{q} \gamma_{\{{\mu}_1}\partial_{{\mu}_2}....\partial_{{\mu}_s\}} q \nonumber \\
&& G^{\rho}_{\{{\mu_1}}\partial_{{\mu}_2}....\partial_{{\mu}_{s-1}}G_{{\mu}_s\} \rho}.
\end{eqnarray}
These operators are of dimension $\Delta = s+2$. Under conformal transformations,
operators scale based on their twist $\Delta-s$. Thus a constant bulk field, representing
the vev of such an operator, should vary as $\phi \sim z^2$. The source for the operator
$\mathcal{J}_{\Delta}$ has dimension $4-\Delta=2-s$ and so it sources a field which varies
like $\phi \sim z^{2-2s}$. This is the same AdS zero mode solution we derived by symmetry
arguments above.

\begin{figure}[htbp]
\begin{center}
\epsfig{file=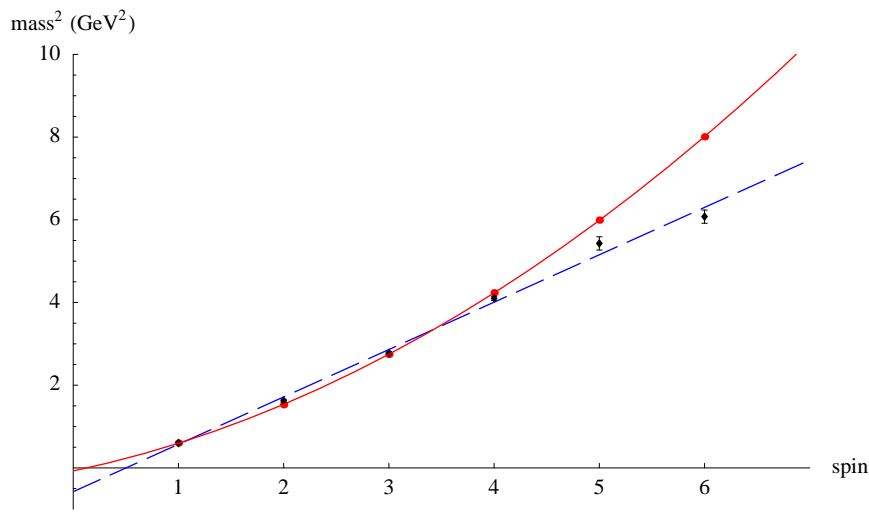,height=3in}
\caption{Squared masses from AdS/QCD (red) verses experiment (blue), 
as a function of spin. Linear (Regge) fit
to experimental values and quadratic fit to AdS/QCD values are shown.
The errors shown are due to the widths of the resonances.}
\end{center}
\label{figspins}
\end{figure}

The normalizable solutions to \eqref{spinkk} are $z^s\mathcal{J}_s(mz)$.  Requiring Neumann 
boundary conditions in the IR, we find that the mass of the lightest
spin-$s$ particle is given by the first zero of $\mathcal{J}_{s - 1}$.
This leads to the following predictions for spin-$s$ resonances in QCD
\begin{equation}
  \begin{array}{|l|l|l|l|l|l|l|}
    \hline
    \mathrm{Particle} & \rho & f_2 & \omega_3 & f_4 & \rho_5 & f_6\\
    \hline
    \mathrm{Experiment} ( \mathrm{MeV} ) & 775.8 & 1275 \pm 1 & 1667 \pm 4 & 2025
    \pm 8 & 2330 \pm 35 & 2465 \pm 50\\
    \hline
    \mathrm{AdS/QCD (MeV)} & -- & 1236 & 1656 & 2058 &
    2448 & 2829\\
    \hline 
  \end{array} \nonumber
\end{equation}
Note that we have neglected the effects of quark masses 
and so we are not sensitive to isospin.
In Figure~\ref{figspins}, we plot the squared masses from this table as
a function of spin. We also display the linear fit to the experimental values,
which is the Regge trajectory, and the quadratic curve through the AdS/QCD
points. Note that the AdS/QCD results take only $m_\rho$ from data, but
provide a good estimate of both the intercept and slope of the Regge
trajectory.
Although the linear fit is an good match to data,
the quadratic fit from AdS/QCD seems to produce deviations at low
energy in the right direction. It is expected that at higher energy, stringy
effects are dominant, and we should merge smoothly back into the linear regime.

We might try to calculate the decay rates of these particles as well. There
are two main impediments. First, none of the decay modes are well measured.
Second, we do not know how to write down interactions for higher spin fields.
For the spin-2 case, we used general coordinate invariance to guess the $f_2$
couplings. But for higher spin fields, it
is impossible to find a consistent set of interactions (there can be no higher
spin conserved current), 
at least in flat space\footnote{See~\cite{Vasiliev:2003ev} for the
intriguing possibility of constructing consistent interacting theories of
higher spin fields in AdS.}. Yet we know massive higher spin fields exist in
QCD, and we can try to guess these interactions, or perhaps compute them using 
the holographic map.

\section{Conclusions\label{SecConc}}
In this paper, we have shown that despite obvious challenges, the
AdS/CFT correspondence produces some remarkably accurate predictions about
QCD. We saw that the mass for the $f_2$ meson and the rate 
for $f_2 \rightarrow \gamma \gamma$ are in fantastic agreement with experiment.
The $f_2 \rightarrow \gamma \gamma$ prediction is particularly satisfying,
because it is difficult to approach this decay through more traditional methods.
Moreover, we have demonstrated that higher-dimension bulk operators do not affect
this decay rate, and so our results are trustworthy. In contrast, we also
calculated the rate for $f_2 \rightarrow \pi \pi$, which was off by a factor
of 4. This rate is sensitive to higher order corrections, and to our
representation of chiral symmetry breaking in AdS/CFT.
Thus our results are predictions of a self-consistent effective field
theory which matches remarkably well to QCD at leading order.

We have also shown that the naive expectation from AdS/CFT, that the lightest 
state of a given spin is captured by the lowest dimension operator with that spin, 
seems to be favored by data.  This indicates that not only is AdS/QCD capable of 
quantitative success, but also that it may reveal interesting connections between QCD 
and its dual. 
For example, we have seen that the $f_2$ makes an awful 4D graviton.
In contrast, on an AdS$_5$ background, general coordinate invariance
allowed us to predict the $f_2$ coupling to photons correctly, making it
a natural 5D graviton KK mode.

\section*{Acknowledgements}
We would like to thank T.~Bhattacharya, M.~Chanowitz, and M.~Suzuki 
for illuminating conversations about meson physics. 
EK and MS thank the Aspen Center of Physics 
for its hospitality during the completion of part of this work.

\end{document}